\documentclass[12pt,preprint]{aastex}

\newcommand\mcnd{\multicolumn{2}{c}{~\nodata}}
\newcommand\mmcnd{\multicolumn{3}{c}{~\nodata}}

\shorttitle{Chemical Composition of NGC~6822}
\shortauthors{Peimbert et al.}

\received{}
\begin{document}

\title{Chemical Composition of Two \ion{H}{2} Regions in NGC~6822 Based on
VLT Spectroscopy
\footnotemark{}}

\author {Antonio Peimbert}
\affil {Instituto de Astronom\'\i a, UNAM,
Apdo. Postal 70-264, M\'exico 04510 D.F., Mexico} 
\email{antonio@astroscu.unam.mx}

\author{Manuel Peimbert}
\affil{Instituto de Astronom\'\i a, UNAM, 
Apdo. Postal 70-264, M\'exico 04510 D.F., Mexico}
\email{peimbert@astroscu.unam.mx}
   
\and 

\author{Mar\'{\i}a Teresa Ruiz}
\affil{Departamento de Astronom\'{\i}a, Universidad de Chile,
Casilla Postal 36D, Santiago de Chile, Chile}
\email{mtruiz@das.uchile.cl}

\begin{abstract}

We present long slit spectrophotometry of regions V and X of the local
group irregular galaxy NGC 6822.  The data consist of VLT FORS
observations in the 3450 to 7500 \AA\ range.  We have obtained
electron temperatures and densities using different line intensity
ratios. We have derived the He, C, and O abundances relative to H
based on recombination lines, the abundance ratios among these
elements are almost independent of the temperature structure of the
nebulae. We have also determined the N, O, Ne, S, Cl, and Ar
abundances based on collisionally excited lines, the ratios of these
abundances relative to that of H depend strongly on the temperature
structure of the nebulae. The chemical composition of NGC 6822 V is
compared with those of the Sun, the Orion nebula, NGC 346 in the SMC,
and 30 Doradus in the LMC. The O/H value derived from recombination
lines is in good agreement with the value derived by \citet{ven01}
from two A type supergiants in NGC 6822.

\end {abstract}


\footnotetext{Based on observations collected at the European Southern
Observatory, Chile, proposal number ESO 69.C-0203(A).}

\section{Introduction}

The main aim of this paper is to make a new determination of the
chemical abundances of the two brightest \ion{H}{2} regions in NGC 6822:
regions V and X \citep{hub25}.  We include the following improvements
over previous determinations: the consideration of the temperature
inhomogeneities that affects the helium and heavy elements abundance
determinations, the derivation of the O and C abundances from
recombination line intensities, the consideration of the collisional
excitation of the triplet \ion{He}{1} lines from the $2^3$S level by
determining the helium abundance from many line intensity ratios, and
the study of the $2^3$S level optical depth effects on the intensity
of the triplet lines by observing a large number of singlet and
triplet lines of \ion{He}{1}.

We are interested in three applications based on the abundance
determinations: the determination of $t^2$, the comparison of the
nebular abundances derived in this paper with the stellar abundances
of supergiant stars in NGC 6822 derived elsewhere
\citep{mus99,ven01,ven02,ven04}, and to provide accurate abundances
for galactic chemical evolution models of this object 
\citep[e. g.][]{car05a}.

NGC 6822 is an irregular galaxy member of the local group particularly 
suited for chemical evolution models because its star formation history 
is well known \citep{wyd01},
and apparently has not been affected by tidal effects, therefore its
chemical composition might permit to decide if outflows to the
intergalactic medium have occurred in this galaxy.

The importance of outflows to the intergalactic medium from nearby
irregular galaxies depends on many factors, like their total mass,
the distribution in time and space of their star formation,
and tidal effects \citep[e.g.][and references 
therein]{leg01, ten03, mar03, fra04}.

NGC 6822 apparently has not
been affected by tidal effects from the main galaxies of the local
group: the Milky Way and M31 \citep{saw05}.
NGC 6822 is located
at 495 kpc from our Galaxy and is moving away from it at a radial velocity of
44 km s$^{-1}$ \citep{tri00}. NGC 6822 is also located at 880 kpc from M31, 
and it is separated from M31 by more than 90 degrees in the sky \citep{tri00}.
 
In sections 2 and 3 the observations and the reduction procedure are
described. In section 4 temperatures and densities are derived from
four and three different intensity ratios respectively; also in this
section, the mean square temperature fluctuation, $t^2$, is determined
from the \ion{O}{2}/[\ion{O}{3}] line intensity ratios and from the
difference between $T$(\ion{He}{1}) and $T$([\ion{O}{3}+\ion{O}{2}]).
In section 5 we determine ionic abundances based on recombination
lines that are almost independent of the temperature structure, we
determine also ionic abundances based on ratios of collisionally
excited lines to recombination lines that do depend on the temperature
structure of the nebula. In section 6 we determine the total
abundances.  In sections 7 and 8 we present the discussion and the
conclusions.

\section{Observations}

The observations were obtained with the Focal Reducer Low Dispersion
Spectrograph 1, FORS1, at the VLT Melipal Telescope in Chile. We used
three grisms: GRIS$_-$600B+12, GRIS$_-$600R+14 with filter GG435, and
GRIS$_-$300V with filter GG375 (see Table~\ref{tobs}).

The slit was oriented almost east-west (position angle 91$^{\rm o}$) to observe
the brightest regions of NGC 6822 V and of 6822 X simultaneously. The linear
atmospheric dispersion corrector, LACD, was used to keep the same observed
region within the slit regardless of the air mass value. The slit width was set
to 0.51" and the slit length was 410". The aperture extractions were made for an
area of 22.4" $\times$ 0.51" for region V and an area of 24.6" $\times$ 0.51"
for region X, covering the brightest parts of both regions (see Figure 1).  The
resolution for the emission lines observed with the blue grism is given by
$\Delta\lambda \sim \lambda / 1300$, with the red grism is given by
$\Delta\lambda \sim \lambda / 1700$, and with the low resolution grism is given
by $\Delta\lambda \sim \lambda / 700$. The average seeing during the
observations amounted to 0.8".

The spectra were reduced using IRAF\footnotemark{} reduction packages,
following the standard procedure of bias subtraction, aperture
extraction, flatfielding, wavelength calibration and flux
calibration. For flux calibration the standard stars LTT~2415,
LTT~7389, LTT~7987 and EG~21 were used \citep{ham92,ham94}.  The
observed spectra are presented in Figures 2 and 3.

\footnotetext{IRAF is distributed by NOAO, which is operated by AURA,
under cooperative agreement with NSF.}

\section{Line Intensities, Reddening Correction, and Radial Velocities}

Line intensities were measured integrating all the flux in the line
between two given limits and over a local continuum estimated by
eye. In the few cases of line-blending, the line flux of each
individual line was derived from a multiple Gaussian profile fit
procedure. All these measurements were carried out with the {\tt
splot} task of the IRAF package.

The reddening coefficients, $C$(H$\beta$)'s, were determined by
fitting the observed $I$(H$\beta$)/$I$(H Balmer lines) ratios to the
theoretical ones computed by \citet{sto95} for $T_e$ = 10,000 K and
$N_e$ = 100 cm$^{-3}$, (see below) and assuming the extinction law of
\citet{sea79}.

Table~\ref{tlines} presents the emission line intensities of the NGC~6822
\ion{H}{2} regions.  The first two columns include the adopted laboratory
wavelength, $\lambda$, and the identification for each line. The third and
fourth columns include the observed flux relative to H$\beta$, $F(\lambda$), and
the flux corrected for reddening relative to H$\beta$, $I(\lambda$), for region
V. The last two columns include the same information as the previous two but for
region X. To combine all the line intensities, from the three different
instrumental settings, on the same scale, we multiplied the intensities in each
setting by a correction factor obtained from the lines present in more than one
setting.  The errors were estimated by comparing all the measured H line
intensities, with those predicted by the computations of \citet{sto95} and by
assuming that the signal to noise increases as the square root of the measured
flux. These estimates are in agreement with the differences found when comparing
the line fluxes observed in different exposures.

\section{Physical Conditions}

\subsection{Temperatures and densities}

The temperatures and densities presented in Table~\ref{tdat} were derived
from the line intensities presented in Table~\ref{tlines}. The
determinations were carried out based on the {\tt temden} IRAF subroutine; this
subroutine models a 5-, 6-, or 8-level ion to derive these quantities.

To compute $T$([\ion{O}{2}]), the contribution to the intensities of the
$\lambda\lambda$ 7319, 7320, 7331, and 7332 [\ion{O}{2}] lines due to
recombination was taken into account based on the following equation:
\begin{equation}
I_R(7319+7320+7331+7332)/I({\rm H\beta})
= 9.36(T/10^4)^{0.44} \times {\rm{O}}^{++}/{\rm{H}}^+
\end{equation}
\citep[see][]{liu00}.

Similarly,
to compute $T$([\ion{N}{2}]), the contribution to the intensity of the
$\lambda$ 5755 [\ion{N}{2}] line due to recombination was taken into account
based on the following equation:
\begin{equation}
I_R(5755)/I({\rm H\beta})
= 3.19(T/10^4)^{0.30} \times {\rm{N}}^{++}/{\rm{H}}^+
\end{equation}
\citep[see][]{liu00}. 
 
\subsection{Temperature variations}

To derive the ionic abundance ratios the average temperature, $T_0$, and the
mean square temperature fluctuation, $t^2$, were used. These quantities are
given by
\begin{equation}
T_0 (N_e, N_i) = \frac{\int T_e({\bf r}) N_e({\bf r}) N_i({\bf r}) dV}
{\int N_e({\bf r}) N_i({\bf r}) dV}
\end{equation}
and
\begin{equation}
t^2 = \frac{\int (T_e - T_0)^2 N_e N_i dV}{T_0^2 \int N_e N_i dV},
\end{equation}
respectively, where $N_e$ and $N_i$ are the electron and the ion densities of
the observed emission line and $V$ is the observed volume \citep{pei67}.

To determine $T_0$ and $t^2$ we need two different methods to derive $T_e$: one
that weighs preferentially the high temperature regions and one that weighs
preferentially the low temperature regions \citep{pei67}. In this paper we have
used the temperature derived from the ratio of the [\ion{O}{3}] $\lambda\lambda$
4363, 5007 lines, $T_{(4363/5007)}$, that is given by
\begin{equation}
T_{(4363/5007)} = T_0 \left[ 1 + {\frac{1}{2}}\left({\frac{91300}{T_0}} - 3
\right) t^2\right]
\end{equation}
and the temperature derived from the ratio of the recombination lines of
multiplet 1 of \ion{O}{2} to the collisionally excited lines of [\ion{O}{3}]
that is given by \citep[see][equations 8-12]{pei04}
\begin{equation}
T_{({\rm O\,\scriptscriptstyle II}\,rec /
{\rm O\,\scriptscriptstyle III}\,coll)} =
T_{(4649/5007)} = f_1(T_0,t^2).
\end{equation}

Using the O recombination lines of Region V, based on these equations, we obtain
$T_0 = 10100$~K and $t^2 = 0.092 \pm 0.026$. Since in this object most of the
oxygen is twice ionized (see section 5.3), this $t^2$ value is representative
for the whole \ion{H}{2} region.

It is also possible to derive the $t^2$ value from the analysis of the 
helium lines (see section 5.1). The resulting values are: $t^2 = 0.060 \pm
0.026$ and $0.056  \pm 0.045$ for regions V and X respectively.
For region V we combine the oxygen and the helium determinations and adopt
$t^2 = 0.076 \pm 0.018$; for region X we simply adopt the $t^2$ helium determination.
The $t^2$ values derived for regions V and X are somewhat larger than those derived
for Galactic \ion{H}{2} regions, that typically are in the 0.03 to 0.04 range
\citep{est05} but are similar to those derived in giant extragalactic
\ion{H}{2} regions \citep{est02,pea03}.

\section{Ionic Chemical Abundances}

\subsection{Helium ionic abundances}

To obtain He$^+$/H$^+$ values we need a set of effective recombination
coefficients for the He and H lines, the contribution due to collisional
excitation to the helium line intensities, and an estimate of the optical depth
effects for the helium lines. The recombination coefficients used were those by
\citet{sto95} for H, and those by \citet{smi96} and \citet{ben99} for He. The
collisional contribution was estimated from \citet{saw93} and
\citet{kin95}. The optical depth effects in the triplet lines were estimated
from the computations by \citet*{ben02}.

Before using the helium lines, much in the same way as we do for hydrogen lines,
we need to correct them for underlying absorption and, in the cases of
$\lambda$~3889 and $\lambda$~4713, for blends. To correct the blue lines,
$\lambda < 5000$~\AA, for underlying absorption we used the values determined by
\citet{gon99}, for the redder lines we used values determined by Cervi\~no
M. (private communication) based on the paper by \citet{gon05}. It should be
noted that the underlying absorption in the helium lines scales, in the same way
as it does in the hydrogen lines, as a fraction of the correction to
H$\beta$. After discarding the lines that could not be easily unblended or those
for which there is no accurate atomic data, the remaining lines are presented in
Table~\ref{tmlm}.

We have many measured helium lines, each of them with a different dependence on
temperature and density. In principle one can find \ion{He}{1} line ratios that
will allow to measure temperature or density. In practice the dependence of each
ratio is weak, also each ratio depends simultaneously on $T_0$, $t^2$, $N_e$,
and $\tau_{3889}$; making the determinations, obtained from any one ratio, to
have large error bars. In order to optimize our data we used a maximum
likelihood method (MLM) to search for the physical and chemical conditions
($T_0$, $t^2$, $\tau_{3889}$, and He$^+$/H$^+$) that would give us the best
simultaneous fit to all the measured lines \citep[see][]{pei00}.

For these objects the MLM can not determine the electron density with high
enough accuracy to be useful; therefore we have adopted the electron densities
derived from collisionally excited lines. Based on Table~\ref{tdat} we adopted
$N_e=175\pm30$~cm$^{-3}$ and $N_e=30\pm30$~cm$^{-3}$, for regions V and X
respectively, to help with the determinations.

To help break the degeneracy on $T_0$ and $t^2$ we need to use an additional
temperature; since $T$(\ion{He}{1}) weighs preferentially the low temperature
regions, we used a temperature that weighs preferentially the high temperature
regions (see section 4.2). In order for this temperature to be representative of
the whole region we used a weighted average of $T$([\ion{O}{2}]) and
$T$([\ion{O}{3}]) \citep[see][]{pea02}: for region V we used
$T$([\ion{O}{3}+\ion{O}{2}])$=12000$~K and for region X we
used$T$([\ion{O}{3}+\ion{O}{2}])$=12250$~K.

{From} these line intensities, densities, temperatures and the MLM we obtained
the He abundances presented in Table~\ref{trionic}, a $t^2 = 0.060 \pm 0.026$
for region V, and a $t^2 = 0.056 \pm 0.045$ for region X.

\subsection{C and O ionic abundances from recombination lines}

The C$^{++}$ abundance was derived from the $\lambda 4267$ \AA~line of
\ion{C}{2} and the effective recombination coefficients computed by
\citet{dav00} for Case A and $T = 10,000$~K.
We only observe the $\lambda 4267$ \AA~line in region~V, the expected
flux of this
line for region~X would produce a S/N$\approx$1, making it
impossible to measure.

The O$^{++}$ abundance was derived from multiplet 1 of \ion{O}{2}
\citep{pei93,sto94}.
The multiplet consists of 8 lines and the sum of their intensities,
$I$(sum), normalized to $I$(H$\beta$), is independent of the electron
density.  On the other hand the normalized intensity of each of the
eight lines does depend on the electron density \citep{rui03}. It is
rarely possible to measure all the lines of this multiplet, and
frequently it is necessary to estimate the intensities of the
unobserved (or blended) lines.

As with \ion{C}{2}, we only observe \ion{O}{2} recombination lines on
region~V. Of the 8 lines of multiplet 1 we only detect 4; which, due to
the dispersion of our observations, are blended into 2 pairs
$\lambda \lambda$ 4639+42 and 4649+51 \AA. 

\citet{rui03} found that, for typical \ion{H}{2} region densities,
the relative intensities of the lines of the multiplet deviate from 
the LTE computation predictions. In order to determine what 
fraction of the intensity of the whole multiplet is emitted in
these 4 lines it is necessary to determine the density dependence 
of each of these lines. Ruiz et al. present the density dependence of
$I$(4649)/$I$(sum). In order to determine the density relations, 
for all the lines of the multiplet, in Figures \ref{f4651},
\ref{f4639}, \ref{f4642}, and \ref{f4649} we present plots of
the data listed by Ruiz et al. in their Table 8 for the four sets of
lines of multiplet 1 that originate in a given upper level (note that our
Figure~\ref{f4649} is the same as Figure 2 by Ruiz et al.). 
From Figures \ref{f4651} to \ref{f4649} 
it can be seen that the intensity of the lines that 
originate in the 3p~$^4$D$^0_{3/2}$ and 3p $^4$D$^0_{1/2}$ energy levels 
decreases with increasing local density, while the intensity of those lines 
that originate in the 3p~$^4$D$^0_{7/2}$ increases with increasing density. 
This is due to the effect 
of collisional redistribution that increases the population of the high 
statistical weight levels at the expense of the low statistical weight ones. 
Note that the intensity of the lines that originate from the 3p~$^4$D$^0_{5/2}$ 
energy level depend weakly on the electron density.

Based on Figures \ref{f4651} to \ref{f4649} and a relationship of the type
\begin{equation}
\left[ \frac{I(line)}{I({\rm sum})} \right]_{obs} =
\left[ \frac{I(line)}{I({\rm sum})} \right]_{LTE} +
\frac{ \left[ \frac{I(line)}{I({\rm sum})} \right]_{cas} -
       \left[ \frac{I(line)}{I({\rm sum})} \right]_{LTE} }
     { \left[ 1 + \frac{N_e({\rm FL})}{N_e({\rm crit})} \right]},
\label{ecurve0}
\end{equation}
we  have obtained the following equations:
\begin{equation}
\left[ \frac{I(4651+74)}{I({\rm sum})} \right]_{obs} =
0.101 + \frac{0.144}{ \left[ 1 + N_e({\rm FL})/1325 \right] },
\label{e4651}
\end{equation}
\begin{equation}
\left[ \frac{I(4639+62+96)}{I({\rm sum})} \right]_{obs} =
0.201 + \frac{0.205}{ \left[ 1 + N_e({\rm FL})/1325 \right] },
\label{e4639}
\end{equation}
\begin{equation}
\left[ \frac{I(4642+76)}{I({\rm sum})} \right]_{obs} =
0.301 - \frac{0.057}{ \left[ 1 + N_e({\rm FL})/1325 \right] },
\label{e4642}
\end{equation}
and
\begin{equation}
\left[ \frac{I(4649)}{I({\rm sum})} \right]_{obs} =
0.397 - \frac{0.292}{ \left[ 1 + N_e({\rm FL})/1325 \right] },
\label{e4649}
\end{equation}
where the first term on the right hand side corresponds to the LTE ratio
presented by \citet{wie96} and the second term takes into account the 
deviation from LTE. The intensity of the lines originating
from the same upper level is constant and depends only on the ratio of
the Einstein A coefficients for each level; thus $I(4651)$:$I(4674)$ is
0.844:0.156, $I(4639)$:$I(4662)$:$I(4696)$ is 0.455:0.506:0.039, and
$I(4642)$:$I(4676)$ is 0.742:0.258; since $\lambda 4649$ \AA~is the only
line originating from the  3p $^4$D$^0_{7/2}$ upper level it will 
contain all the photons originating from this level.

Based on equations \ref{e4651} to \ref{e4649}, the relative intensities of the
lines originating from the same upper level, and the \ion{O}{2} lines presented
in Table~\ref{tlines} we have determined the $I$(sum) value for multiplet 1
which amounts to $I$(sum)/ $I$(H$\beta$)=0.0023 . We have derived the O$^{++}$
abundance presented in Table~\ref{trionic} from $I$(sum) and the effective
recombination coefficient for multiplet 1 computed by \citet{sto94} under the
assumption of Case B for $T_e = 10,000$ K and $N_e$ = 100 cm$^{-3}$.

\subsection{Ionic abundances from collisionally excited lines}

The values presented in Table~\ref{tceionic} for $t^2 = 0.00$ were derived with
the IRAF task {\tt abund}, using only the low- and medium-ionization zones.
{\tt abund} requires as inputs a temperature and a density for each zone as well
as the intensities of the observed collisionally excited lines relative to
H$\beta$. These values are combined with a model of a 5-, 6-, or 8-level ion to
determine the abundances. The low and medium ionization zones of {\tt abund}
correspond to the low and high ionization zones of this paper. For Region V we
used $T_{low}=13000\;$K, $T_{high}=11900\;$K, and
$N_{e\;low}=N_{e\;high}=175\;$cm$^{-3}$; and for Region X we used
$T_{low}=13300\;$K, $T_{high}=12000\;$K, and
$N_{e\;low}=N_{e\;high}=30\;$cm$^{-3}$.

To derive the abundances for $t^2 > 0.00$ we used the abundances for
$t^2 = 0.00$ corrected by the formulation for $t^2 > 0.00$ presented
by \citet{pei69} \citep[see also][]{pei04}. To derive abundances for
other $t^2$ values it is possible to interpolate or to extrapolate the
values presented in Table~\ref{tceionic}.

\section{Total Abundances}

To obtain the C and N gaseous abundances the following equations were adopted
\begin{equation}
\label{ecarbon}
\frac{N(\rm C)}{N(\rm H)} = ICF({\rm C}) \frac{N({\rm C^{++}})}{N(\rm H^+)}
\end{equation}
and
\begin{equation}
\label{enitrogen}
\frac{N(\rm N)}{N(\rm H)} = ICF({\rm N}) \frac{N({\rm N^{+}})}{N(\rm H^+)},
\end{equation}
the $ICF$(C) value was obtained from
\citet{gar95} and amounts to 0.07~dex. The $ICF$(N) value was obtained from the models by \citet{moo04} 
and amounts to 1.13 dex for region V and to 0.91 dex for region X. Note that 
the $ICF$(N) value predicted by \citet{pei69} is given by
$N$(O)/$N$(O${^+}$) and amounts to 0.97 dex for region V and to 0.75 dex for
region X; the $ICF$(N) formula by Peimbert and Costero is a very good
approximation for \ion{H}{2} regions with low degree of ionization but
becomes only fair for \ion{H}{2} regions with high degree of ionization.
The $ICF$(N) for models in which a large fraction of the ionizing photons
escape from the \ion{H}{2} region becomes even larger than those predicted
by the models by \citet{moo04}, see for example the results for NGC~346
obtained by \citet{rel02}.

The gaseous abundances for O and Ne were obtained from the following
equations \citep{pei69}
\begin{equation}
 \frac{N(\rm O)}{N(\rm H)} =
             \frac{N({\rm O^+})+N(\rm O^{++})}{N(\rm H^+)}
\end{equation}
and
\begin{equation}
\frac{N(\rm Ne)}{N(\rm H)} =
             \left( \frac{N({\rm O^+})+N(\rm O^{++})}{N(\rm O^{++})} \right)
             \frac{N(\rm Ne^{++})}{N(\rm H^+)}.
\end{equation}

The gaseous abundances of S, Cl, and Ar, were obtained from the following
equations:
\begin{equation}
\label{esulphur}
\frac{N(\rm S)}{N(\rm H)} = ICF({\rm S}) \frac{N({\rm S^+}) + N(\rm
S^{++})}{N(\rm H^+)},
\end{equation}
\begin{equation}
\label{echlorine}
\frac{N(\rm Cl)}{N(\rm H)} = ICF({\rm Cl}) \frac{N({\rm Cl^{++}})}{N(\rm H^+)},
\end{equation}
and
\begin{equation}
\label{eargon}
\frac{N(\rm Ar)}{N(\rm H)} = ICF({\rm Ar}) \frac{N({\rm Ar^{++}}) + N(\rm
Ar^{3+})}{N(\rm H^+)}.
\end{equation}
The $ICF$(S) values were estimated from the models by \citet{gar89} and amount to
0.22 dex for region V and to 0.10 dex for region X. The $ICF$(Cl) values amount to
0.05 dex for regions V and X and were obtained by averaging the observed 
$N$(Cl$^+$ + Cl$^{++}$ + Cl$^{+3}$)/ $N$(Cl$^{++}$) values obtained for 
Orion, 30 Doradus, and NGC 3576 \citep{est04,pea03,gar04}. The ionization
correction factor due to the Ar$^{+}$ fraction was estimated from
$ICF$(Ar) = 1/[1 - O$^+$/O] \citep{liu00} and amounts to 0.05 dex and 0.09 dex
for regions V and X respectively.

Based on the previous considerations, we present the total gaseous abundances 
of regions V and X in Tables \ref{tabundV} and \ref{tabundX} respectively. The
errors for region X are larger than those for region V because the
brightness of region X is about three times smaller than that of
region V. Within the errors the abundances of both regions are similar.
We need observations of higher quality, than those presented in this paper,
to establish the presence of abundance variations  among \ion{H}{2} regions 
in NGC 6822.

In  Table~\ref{tta} we present the adopted total abundances for NGC~6822.
To obtain the total O and C abundances we have to add a correction to the
gaseous abundance due to the presence of dust. Following \citet{est98} this
corrections amount to 0.08~dex for O and 0.10~dex for C.

We have also computed the hydrogen, helium, and heavy elements by mass
for NGC 6822 V and amount to: $X$ = 0.7501, $Y$ = 0.2433, and $Z$ =
0.0066. The $Z$ value was computed from the C, N, O, Ne, and Fe
abundances \citep[see Table~\ref{tta}, where the Fe abundance comes
from stellar data by][]{ven01} and assuming that they constitute 83.3$\%$
of the total $Z$ value. This fraction was obtained by assuming that
all the other heavy elements in NGC 6822 present the same abundances
relative to O than in the Sun as presented by \citet{asp05}.

\section{Discussion}

\subsection{Comparison with other nebular abundance determinations}

There have been seven O/H determinations for Hubble V carried out by
\citet{smi75, leq79, tal80, pei70, pag80, ski89} and \citet{hid01}
that amounted to 12 + log O/H = 8.45, 8.28, 8.20, 8.10, 8.19, 8.20 and
8.10 respectively; these determinations were made based on the
$T$(\ion{O}{3}) temperature and under the assumption that $t^2 =
0.00$. Statistical errors are typically about 0.1 dex, so to a first
approximation there is good agreement among the seven
determinations. A second look reveals a systematic difference, the
first three determinations were made with the image intensified
dissector scanner, IIDS, and yield an average value of 8.31 dex, while
the other four were made with other detectors and yield an average
value of 8.15 dex, the difference is real and is mainly due to the
non-linearity of the IIDS detector \citep{pei87}.

In Tables \ref{tabundV} and \ref{tabundX} we compare our determinations with
those of \citet{leq79} and \citet{hid01} for $t^2 = 0.00$. As expected our
values for O/H and Ne/H are in better agreement with those by \citet{hid01}. As
mentioned above the reason is the non linearity of the IIDS detector that
increases the difference between weak lines and strong lines, thus yielding
lower temperatures and consequently larger O/H and Ne/H values. For Region~V
this non linearity also yields lower He$^+$/H$^+$ values , because the helium
line intensities are weaker than the H line intensities. For region~X the
He$^+$/H$^+$ differences are not significative because the observational errors
are larger than the differences predicted by the non-linearity. The smaller
errors for region~V than for region~X are due to the higher luminosity of
region~V, as can be seen in Table~\ref{tlines}. Due to the smaller errors in the
line intensities of region~V, we took the abundances of this region as
representatives of the whole galaxy.

\subsection{Comparison with stellar abundance determinations}

The stars and \ion{H}{2} regions of NGC 6822 have reliable O/H
determinations that can be compared. In Table~\ref{toxy} we present
the O/H abundances derived in this paper for Hubble V with those
derived by \citet{ven01} for two A type supergiants; the A supergiants
were formed a few million years ago, therefore we expect them to have
the same abundances as the \ion{H}{2} regions. We find excellent
agreement between the stellar O/H values and the ones found using
recombination lines; alternatively, the agreement with the
values derived using collisionally excited lines ($t^2 = 0.00$) is poor.
Also, based on three B supergiant stars in NGC 6822 \citet{mus99} find
a Fe/H of -0.5 $\pm$ 0.2 dex relative to the Sun in excellent
agreement with the results of \citet{ven01} for two A type
supergiants.

\subsection{Comparison with the Magellanic Clouds, the Orion nebula, 
and the Sun}

Also in Table~\ref{tta} we present the Orion and the solar abundances
for comparison. For Orion: Fe comes from \citet{cun94} based on Orion
B stars, and all the other elements from \citet{est04}. For the Sun:
He comes from \citet{chr98}, and all the other elements from
\citet{asp05}, the solar abundances are the photospheric ones with the
exception of He that corresponds to the initial He abundance and of Cl
that comes from meteoritic data.

We consider that the Orion nebula is more representative of the
present chemical composition of the solar vicinity than the Sun for
two main reasons: a) the Sun appears deficient by roughly 0.1 dex in
O, Si, Ca, Sc, Ti, Y, Ce, Nd and Eu, compared with its immediate
neighbors with similar iron abundances \citep{all04}, the probable
reason for this difference is that the Sun is somewhat older than the
comparison stars; and b) all the chemical evolution models of the Galaxy
predict a steady increase of the O/H ratio in the solar vicinity with
time, for example the chemical evolution models of the solar vicinity
presented by \citet{car03} and \citet{car05b} indicate that the O/H
value in the solar vicinity has increased by 0.13~dex since the Sun
was formed.


There are two independent methods to determine the O/H ratio in the ISM of the
solar vicinity: a) From the solar 12 + log O/H ratio by \citet{asp05} that
amounts to 8.66 and taking into account the increase of the O/H ratio due to
galactic chemical evolution since the Sun was formed, that according to state of
the art chemical evolution models of the Galaxy amounts to 0.13 dex
\citep[e. g.][]{car05b}, we obtain a value of 8.79 dex; b) The O/H value derived
by \citet{est05}, based on the galactic gradient determined from the \ion{O}{2}
recombination lines observed in \ion{H}{2} regions, that amounts to 8.79 dex in
perfect agreement with the value based on the solar abundance and the chemical
evolution of the Galaxy.

\citet{lee05} have derived a value of 12 + log O/H = 8.45 for the A
supergiant WLM 15 in the Local Group dwarf irregular galaxy WLM, they
have also derived a value of 12 + log O/H = 7.83 for two \ion{H}{2}
regions in the same galaxy based on the $T$(4363/5007) temperature and
under the assumption that $t^2 = 0.00$. The discrepancy between these
two determinations can be reduced by assuming that 20\% of the oxygen
atoms are trapped in dust grains inside \ion{H}{2} regions, as is the
case in the Orion nebula \citep{est98}, and by considering the
presence of spatial temperature fluctuations. A typical correction for
spatial temperature fluctuations based on the recombination lines of
\ion{O}{2} in NGC~6822V and other extragalactic \ion{H}{2} regions
\citep{est02} amounts to about 0.25 dex.

\section{Conclusions}

We show that the ratios of the lines of multiplet 1 of \ion{O}{2} are
not constant, but instead they depend on density. Only at high
densities they coincide with the ratios predicted using LTE. The
same arguments can probably be made for most recombination lines from
other multiplets of heavy elements.

We present a set of empirical equations to determine the O abundances
from recombination lines of multiplet 1 of \ion{O}{2}. These equations
are useful for those cases in which not all the lines of the
multiplet are observed.

We derive for the first time the C/H and O/H abundances of NGC 6822
based on recombination lines. We have also derived the N, O, Ne, S,
Cl, and Ar abundances based on collisionally excited lines for $t^2 =
0.00$ and for $t^2 > 0.00$ (the $t^2$ derived in this paper).

The O/H ratio derived from \ion{O}{2} recombination lines is 0.26 dex higher 
than that derived from [\ion{O}{3}] $\lambda\lambda 4363, 4959, 5007$ collisionally
excited lines under the assumption that $t^2 = 0.00$ (that is, by adopting 
the $T$(4363/5007) temperature).

We have found that, in NGC~6822, the O/H abundance ratio derived from
\ion{O}{2} recombination lines is in very good agreement with the O/H
ratios derived from A-type supergiants, while the O/H abundances
derived from collisionally excited lines are not.  This result is
qualitatively equivalent to that found for the solar vicinity by
\citet{est04} and \citet{car05a}; these authors find that the
\ion{H}{2} region O/H recombination abundances are in agreement with
the O/H solar abundances after considering the O/H enrichment
predicted by a chemical evolution model of the Galaxy, while the
\ion{H}{2} region O/H abundances derived from collisionally excited
lines (assuming $t^2 = 0.00$) are not.

The high N/O ratio places NGC~6822 on the plateau formed by irregular galaxies
in the N/O-O/H diagram, which implies that a large fraction of the N present in
this object is already due to secondary production. From the O/H, C/O, and N/O
values it follows that NGC 6822 is considerably more chemically evolved than the
SMC. Chemical evolution models for NGC 6822 are presented elsewhere
\citep{car05a}.

We are grateful to Leticia Carigi and Pedro Col\'{\i}n for several useful
discussions. We are also grateful to Miguel Cervi\~no for providing us with the
correction for underlying absorption of the \ion{He}{1} red lines. We also
acknowledge several excellent suggestions by Richard B.C. Henry, who was the
referee of this paper, for a careful reading of the manuscript and several
useful suggestions. AP received partial support from DGAPA UNAM (grant IN
118405). MP received partial support from DGAPA UNAM (grant IN 114601). MTR
received partial support from FONDAP(15010003), and Fondecyt(1010404).

\clearpage

\begin{deluxetable}{l@{\hspace{48pt}}r@{--}lc} 
\tablecaption{Journal of Observations
\label{tobs}}
\tablewidth{0pt}
\tablehead{
\colhead{Date} & 
\multicolumn{2}{c}{$\lambda$ (\AA)} &
\colhead{Exp. Time (s)}}
\startdata
2002 Sept  10   & 3450 &5900  &  3$\times$720  \\
2002 Sept  10   & 5250 &7450  &  3$\times$600  \\
2002 Sept  12   & 3850 &7500  &  1$\times$300  \\
\enddata
\end{deluxetable} 

\clearpage

\begin{deluxetable}{llr@{}lr@{}l@{$\pm$}lcr@{}lr@{}l@{$\pm$}l}
\tablecaption{Line Intensities for Regions V and X
\label{tlines}}
\tablewidth{0pt}
\small
\tablehead{
&&& \multicolumn{5}{c}{Region V} &
\multicolumn{5}{c}{Region X} \\
\cline{3-7}\cline{9-13} \\
\colhead{$\lambda$} & \colhead{Id.} &
\multicolumn{2}{c}{$F(\lambda)$\tablenotemark{a}} &
\multicolumn{3}{c}{$I(\lambda)$\tablenotemark{b}} &&
\multicolumn{2}{c}{$F(\lambda)$\tablenotemark{c}} &
\multicolumn{3}{c}{$I(\lambda)$\tablenotemark{d}} }
\startdata
3634   &\ion{He}{1}      &  0.&12&  0.&18& 0.07&& \mcnd &   \mmcnd    \\
3687   &H\,19            &  0.&42&  0.&62& 0.12&& \mcnd &   \mmcnd    \\
3692   &H\,18            &  0.&51&  0.&75& 0.13&& \mcnd &   \mmcnd    \\
3697   &H\,17            &  0.&53&  0.&77& 0.14&& \mcnd &   \mmcnd    \\
3704   &H\,16            &  0.&85&  1.&24& 0.17&&  1.&01&  1.&31& 0.29\\
3712   &H\,15            &  0.&76&  1.&10& 0.16&&  0.&93&  1.&21& 0.26\\
3722   &H\,14            &  1.&14&  1.&64& 0.20&&  1.&12&  1.&44& 0.30\\
3726   &[\ion{O}{2}]     & 26.&90& 38.&90& 1.00&& 46.&80& 60.&40& 2.00\\
3729   &[\ion{O}{2}]     & 37.&80& 54.&60& 1.20&& 66.&90& 86.&30& 2.50\\
3734   &H\,13            &  1.&28&  1.&85& 0.21&&  1.&32&  1.&70& 0.33\\
3750   &H\,12            &  1.&44&  2.&06& 0.22&&  2.&68&  3.&43& 0.47\\
3771   &H\,11            &  2.&02&  2.&87& 0.26&&  2.&32&  2.&95& 0.44\\
3798   &H\,10            &  2.&49&  3.&51& 0.28&&  3.&81&  4.&82& 0.55\\
3820   &\ion{He}{1}      &  0.&42&  0.&58& 0.12&&  0.&91&  1.&15& 0.27\\
3835   &H\,9             &  4.&03&  5.&61& 0.36&&  4.&29&  5.&39& 0.58\\
3869   &[\ion{Ne}{3}]    & 25.&30& 34.&70& 0.90&& 21.&90& 27.&30& 1.30\\
3889   &H\,8+\ion{He}{1} & 11.&60& 16.&10& 0.60&& 13.&90& 17.&10& 1.00\\
3967   &[\ion{Ne}{3}]+H\,7+\ion{He}{1}
                         & 19.&00& 25.&40& 0.70&& 18.&90& 23.&00& 1.20\\
4026   &\ion{He}{1}      &  1.&21&  1.&60& 0.18&&  1.&40&  1.&69& 0.31\\
4069   &[\ion{S}{2}]     &  0.&75&  0.&97& 0.14&& \mcnd &   \mmcnd    \\
4076   &[\ion{S}{2}]     &  0.&17&  0.&22& 0.07&& \mcnd &   \mmcnd    \\
4101   &H$\delta$        & 19.&20& 24.&70& 0.70&& 20.&20& 24.&00& 1.20\\
4144   &\ion{He}{1}      &  0.&19&  0.&24& 0.07&& \mcnd &   \mmcnd    \\
4146+53&\ion{O}{2}+\ion{Ne}{2}
                         &  0.&13&  0.&16& 0.06&& \mcnd &   \mmcnd    \\
4192+96& \ion{O}{2}      &  0.&12&  0.&15& 0.05&& \mcnd &   \mmcnd    \\
4267   &\ion{C}{2}       &  0.&07&  0.&09& 0.02&& \mcnd &   \mmcnd    \\
4340   &H$\gamma$        & 39.&20& 47.&00& 1.00&& 41.&40& 46.&90& 1.70\\
4363   &[\ion{O}{3}]     &  4.&87&  5.&79& 0.33&&  4.&21&  4.&73& 0.52\\
4388   &\ion{He}{1}      &  0.&32&  0.&46& 0.09&& \mcnd &   \mmcnd    \\
4471   &\ion{He}{1}      &  3.&48&  3.&97& 0.27&&  3.&50&  3.&82& 0.45\\
4591+96&\ion{O}{2}       &  0.&06&  0.&07& 0.02&& \mcnd &   \mmcnd    \\
4639+42&\ion{O}{2}       &  0.&10&  0.&11& 0.03&& \mcnd &   \mmcnd    \\
4649+51&\ion{O}{2}       &  0.&06&  0.&06& 0.02&& \mcnd &   \mmcnd    \\
4658   &[\ion{Fe}{3}]    &  0.&23&  0.&24& 0.06&& \mcnd &   \mmcnd    \\
4711   &[\ion{Ar}{4}] + \ion{He}{1}   
                         &  0.&83&  0.&86& 0.12&&  0.&26&  0.&27& 0.11\\
4740   &[\ion{Ar}{4}]    &  0.&31&  0.&32& 0.07&& \mcnd &   \mmcnd    \\
4861   &H$\beta$         &100.&00& 99.&10& 1.30&&100.&00& 99.&10& 2.30\\
4922   &\ion{He}{1}      &  1.&07&  1.&04& 0.12&&  1.&17&  1.&14& 0.23\\
4959   &[\ion{O}{3}]     &183.&00&177.&00& 2.00&&147.&67&144.&00& 3.00\\
5007   &[\ion{O}{3}]     &557.&00&535.&00& 4.00&&439.&00&426.&00& 5.00\\
5016   &\ion{He}{1}      &  2.&55&  2.&40& 0.19&&  2.&77&  2.&65& 0.35\\
5048   &\ion{He}{1}      &  0.&16&  0.&17& 0.05&& \mcnd &   \mmcnd    \\
5270   &[\ion{Fe}{3}]    &  0.&08&  0.&07& 0.03&& \mcnd &   \mmcnd    \\
5517   &[\ion{Cl}{3}]    &  0.&48&  0.&39& 0.07&&  0.&47&  0.&40& 0.13\\
5537   &[\ion{Cl}{3}]    &  0.&39&  0.&31& 0.06&&  0.&29&  0.&25& 0.10\\
5755   &[\ion{N}{2}]     &  0.&22&  0.&16& 0.04&& \mcnd &   \mmcnd    \\
5876   &\ion{He}{1}      & 15.&80& 11.&40& 0.40&& 14.&20& 11.&20& 0.70\\
6151   &\ion{C}{2}       &  0.&08&  0.&05& 0.02&& \mcnd &   \mmcnd    \\
6312   &[\ion{S}{3}]     &  2.&90&  1.&88& 0.14&&  2.&68&  1.&98& 0.27\\
6548   &[\ion{N}{2}]     &  2.&37&  1.&46& 0.12&&  3.&35&  2.&39& 0.29\\
6563   &H$\alpha$        &462.&00&285.&00& 2.00&&401.&00&285.&00& 4.00\\
6584   &[\ion{N}{2}]     &  8.&33&  5.&12& 0.23&& 10.&00&  7.&10& 0.50\\
6678   &\ion{He}{1}      &  5.&25&  3.&17& 0.17&&  4.&20&  2.&94& 0.32\\
6716   &[\ion{S}{2}]     & 11.&20&  6.&70& 0.26&& 14.&80& 10.&30& 0.60\\
6731   &[\ion{S}{2}]     &  8.&42&  5.&02& 0.22&& 10.&30&  7.&17& 0.50\\
6734   &\ion{C}{2}       &  0.&11&  0.&07& 0.03&& \mcnd &   \mmcnd    \\
7065   &\ion{He}{1}      &  4.&42&  2.&50& 0.15&&  3.&34&  2.&23& 0.27\\
7136   &[\ion{Ar}{3}]    & 16.&50&  9.&20& 0.29&& 13.&70&  9.&06& 0.55\\
7281   &\ion{He}{1}      &  1.&21&  0.&66& 0.07&&  0.&75&  0.&49& 0.12\\
7320   &[\ion{O}{2}]     &  3.&22&  1.&74& 0.12&&  3.&36&  2.&18& 0.27\\
7330   &[\ion{O}{2}]     &  2.&29&  1.&24& 0.10&&  2.&97&  1.&92& 0.25\\
7751   &[\ion{Ar}{3}]    &  4.&36&  2.&22& 0.14&& \mcnd &   \mmcnd    \\
8502   &Pa\,16           &  0.&87&  0.&40& 0.06&& \mcnd &   \mmcnd    \\
8542   &Pa\,15           &  1.&28&  0.&59& 0.07&& \mcnd &   \mmcnd    \\
8596   &Pa\,14           &  1.&19&  0.&55& 0.06&& \mcnd &   \mmcnd    \\
8665   &Pa\,13           &  2.&17&  0.&99& 0.09&& \mcnd &   \mmcnd    \\
8750   &Pa\,12           &  2.&16&  0.&97& 0.08&& \mcnd &   \mmcnd    \\
\hline
\multicolumn{2}{l}{log $EW$(H$\beta$)\tablenotemark{e}}&
\multicolumn{5}{c}{225}&&\multicolumn{5}{c}{215}\\
\multicolumn{2}{l}{$C$(H$\beta$)\tablenotemark{f}}     &
\multicolumn{5}{c}{$0.64\pm0.05$}&&\multicolumn{5}{c}{$0.45\pm0.05$}\\
\enddata
\tablenotetext{a} {$F(\lambda)$ is the observed flux in units of 
$100.00=1.015\times10^{-13}$ erg s$^{-1}$ cm$^{-2}$.}
\tablenotetext{b} {$I(\lambda)$ is the reddening corrected flux in units of 
$100.00=4.43\times10^{-13}$ erg s$^{-1}$ cm$^{-2}$.}
\tablenotetext{c} {$F(\lambda)$ is the observed flux in units of 
$100.00=3.73\times10^{-14}$ erg s$^{-1}$ cm$^{-2}$.}
\tablenotetext{d} {$I(\lambda)$ is the reddening corrected flux in units of 
$100.00=1.050\times10^{-13}$ erg s$^{-1}$ cm$^{-2}$.}
\tablenotetext{e} {$EW$(H$\beta$) is the equivalent width in emission 
given in \AA .}
\tablenotetext{f} {$C$(H$\beta$) is the logarithmic reddening correction.}
\end{deluxetable}
\clearpage

\begin{deluxetable}{lr@{}lr@{}l}
\tablecaption{Densities and Temperatures
\label{tdat}}
\tablewidth{0pt}
\tablehead{
 & \multicolumn{2}{c}{Region V} & \multicolumn{2}{c}{Region X} }
\startdata
Densities (cm$^{-3}$)\\ 
\ion{O}{2}   &              $190$&$\pm30$  &               $30$&$\pm30$ \\
\ion{S}{2}   &               $90$&$\pm75$  & \multicolumn{2}{c}{$<100$} \\
\ion{Cl}{3}  & \multicolumn{2}{c}{$<1000$} & \multicolumn{2}{c}{$<1000$}\\
\hline
Temperatures (K)\\ 
\ion{O}{2}   &            $13000$&$\pm1000$&            $13300$&$\pm900$\\
\ion{N}{2}   &            $15500$&$\pm2500$&               \mcnd        \\
\ion{S}{2}   &            $11500$&$\pm2000$&               \mcnd        \\
\ion{O}{3}   &            $11900$&$\pm250$ &            $12000$&$\pm400$\\
\enddata
\end{deluxetable}
\clearpage

\begin{deluxetable}{lr@{}lr@{}l}
\tabletypesize{\small}
\tablecaption{Input line intensity values for the MLM
\label{tmlm}\tablenotemark{a}}
\tablewidth{0pt}
\tablehead{
\colhead{$\lambda$}             &
\multicolumn{2}{c}{Region V}     & \multicolumn{2}{c}{Region X}}
\startdata
3820  \hspace{18pt}   &  \hspace{18pt}0.&77 &  \hspace{18pt}1.&32 \\ 
3889    &  6.&92 &  7.&87 \\
4026    &  1.&90 &  1.&97 \\
4388    &  0.&54 & \mcnd  \\
4471    &  4.&17 &  4.&01 \\
4713    &  0.&41 & \mcnd  \\
4922    &  1.&13 &  1.&23 \\
5048    &  0.&23 & \mcnd  \\
5876    & 11.&50 & 11.&30 \\
6678    &  3.&21 &  2.&99 \\
7065    &  2.&55 &  2.&29 \\
7281    &  0.&68 &  0.&52 \\
\enddata
\tablenotetext{a}{Given in units of $I($H$\beta)=100$.}
\end{deluxetable}

\clearpage
\begin{deluxetable}{lccccc}
\tabletypesize{\small}
\tablecaption{Ionic Abundance Determinations from Recombination
Lines\tablenotemark{a}
\label{trionic}}
\tablewidth{0pt}
\tablehead{
\colhead{Ion}             &
\multicolumn{2}{c}{Region V}     && \multicolumn{2}{c}{Region X}\\
\cline{2-3}  \cline{5-6}
                                  & \colhead{$t^2 = 0.000$}      &
\colhead{$t^2 = 0.076\pm0.018$}  && \colhead{$t^2 = 0.000$}      &
\colhead{$t^2 = 0.054\pm0.045$}       }
\startdata
He$^+$    & 10.922$\pm$0.010 & 10.909$\pm$0.011 && 10.923$\pm$0.017 & 10.916$\pm$0.018 \\
C$^{++}$  &  7.94$\pm$0.11   &   7.94$\pm$0.11  && \nodata          & \nodata          \\
O$^{++}$  &  8.32$\pm$0.09   &   8.32$\pm$0.09  && \nodata          & \nodata          \\  
\enddata
\tablenotetext{a} {In units of $12 +$ Log $N$(X$^{+i}$)/$N$(H),
 gaseous content only.}

\end{deluxetable}
\clearpage
\begin{deluxetable}{lccccc}
\tabletypesize{\small}
\tablecaption{Ionic Abundance Determinations from Collisionally Excited 
Lines\tablenotemark{a}
\label{tceionic}}
\tablewidth{0pt}
\tablehead{
\colhead{Ion}             & 
\multicolumn{2}{c}{Region V}     && \multicolumn{2}{c}{Region X}\\
\cline{2-3}  \cline{5-6}
                                  & \colhead{$t^2 = 0.000$}      &
\colhead{$t^2 = 0.076\pm0.018$}  && \colhead{$t^2 = 0.000$}      &
\colhead{$t^2 = 0.054\pm0.045$}       }
\startdata
N$^+$     & 5.72$\pm$0.08 & 5.92$\pm$0.10 && 5.85$\pm$0.07 & 5.99$\pm$0.13 \\
O$^+$     & 7.11$\pm$0.12 & 7.37$\pm$0.13 && 7.26$\pm$0.12 & 7.44$\pm$0.19 \\
O$^{++}$  & 8.03$\pm$0.03 & 8.29$\pm$0.06 && 7.92$\pm$0.05 & 8.10$\pm$0.16 \\
Ne$^{++}$ & 7.30$\pm$0.03 & 7.58$\pm$0.07 && 7.18$\pm$0.05 & 7.37$\pm$0.17 \\
S$^+$     & 5.17$\pm$0.07 & 5.37$\pm$0.09 && 5.33$\pm$0.06 & 5.47$\pm$0.13 \\
S$^{++}$  & 6.36$\pm$0.04 & 6.55$\pm$0.06 && 6.37$\pm$0.07 & 6.50$\pm$0.13 \\
Cl$^{++}$ & 4.42$\pm$0.05 & 4.66$\pm$0.08 && 4.38$\pm$0.10 & 4.55$\pm$0.17 \\
Ar$^{++}$ & 5.76$\pm$0.02 & 5.97$\pm$0.05 && 5.75$\pm$0.04 & 5.90$\pm$0.14 \\
Ar$^{+3}$ & 4.63$\pm$0.09 & 4.91$\pm$0.11 &&    \nodata    &    \nodata    \\
\enddata
\tablenotetext{a} {In units of $12 +$ Log $N$(X$^{+i}$)/$N$(H),
 gaseous content only.}

\end{deluxetable}
\clearpage

\begin{deluxetable}{ccccc}
\tabletypesize{\small}
\tablecaption{NGC~6822~V Gaseous Abundance Determinations\tablenotemark{a}
\label{tabundV}}
\tablewidth{0pt}
\tablehead{
\colhead{Element}             & 
\multicolumn{2}{c}{This paper}& \colhead{L\tablenotemark{b}} &
\colhead{H\tablenotemark{c}} \\
\cline{2-3}
                              & \colhead{$t^2 = 0.000$}      &
\colhead{$t^2 = 0.076$}       & \colhead{$t^2 = 0.000$}      &
\colhead{$t^2 = 0.000$}       }
\startdata
He\tablenotemark{d}    &   10.922$\pm$0.010 & 10.909$\pm$0.011  &   10.88    &   10.90   \\
C\tablenotemark{d}     &    8.01 $\pm$0.12  &  8.01 $\pm$0.12   & ~\nodata   & ~\nodata  \\ 
N\tablenotemark{e}     &    6.85 $\pm$0.15  &  7.05 $\pm$0.16   &    6.53    &    6.52   \\
O\tablenotemark{d}     &    8.37 $\pm$0.09  &  8.37 $\pm$0.09   & ~\nodata   & ~\nodata  \\
O\tablenotemark{e}     &    8.08 $\pm$0.03  &  8.34 $\pm$0.06\tablenotemark{f}  
                                                                &    8.20    &    8.10   \\
Ne\tablenotemark{e}    &    7.35 $\pm$0.03  &  7.63 $\pm$0.07   &    7.54    &    7.32   \\
S\tablenotemark{e}     &    6.61 $\pm$0.05  &  6.80 $\pm$0.07   & ~\nodata   & ~\nodata  \\
Cl\tablenotemark{e}    &    4.47 $\pm$0.05  &  4.71 $\pm$0.08   & ~\nodata   & ~\nodata  \\
Ar\tablenotemark{e}    &    5.84 $\pm$0.03  &  6.06 $\pm$0.05   & ~\nodata   & ~\nodata  \\
\enddata
\tablenotetext{a} {In units of $12 +$ Log $N$(X)/$N$(H).}
\tablenotetext{b} {\citet{leq79}.}
\tablenotetext{c} {\citet{hid01}.}
\tablenotetext{d} {Recombination lines.}
\tablenotetext{e} {Collisionally excited lines.}
\tablenotetext{f} {This is the adopted value, note that the $t^2$ 
value implicitly includes information from the \ion{O}{2} 
recombination lines.}

\end{deluxetable}
\clearpage
\begin{deluxetable}{ccccc}
\tabletypesize{\small}
\tablecaption{NGC~6822~X Gaseous Abundance Determinations\tablenotemark{a}
\label{tabundX}}
\tablewidth{0pt}
\tablehead{
\colhead{Element}             & 
\multicolumn{2}{c}{This paper}& \colhead{L\tablenotemark{b}} &
\colhead{H\tablenotemark{c}} \\
\cline{2-3}
                              & \colhead{$t^2 = 0.000$}      &
\colhead{$t^2 = 0.054$}       & \colhead{$t^2 = 0.000$}      &
\colhead{$t^2 = 0.000$}       }
\startdata
He\tablenotemark{d}    &   10.923$\pm$0.010 & 10.916$\pm$0.011  &   10.92    &   11.0    \\ 
N\tablenotemark{e}     &    6.76 $\pm$0.16  &  6.90 $\pm$0.22   &    6.50    &    6.4    \\
O\tablenotemark{e}     &    8.01 $\pm$0.05  &  8.19 $\pm$0.16   &    8.27    &    8.12   \\
Ne\tablenotemark{e}    &    7.27 $\pm$0.05  &  7.46 $\pm$0.17   &    7.61    &    7.4    \\
S\tablenotemark{e}     &    6.51 $\pm$0.06  &  6.64 $\pm$0.13   & ~\nodata   & ~\nodata  \\
Cl\tablenotemark{e}    &    4.43 $\pm$0.10  &  4.60 $\pm$0.17   & ~\nodata   & ~\nodata  \\
Ar\tablenotemark{e}    &    5.84 $\pm$0.05  &  5.99 $\pm$0.14   & ~\nodata   & ~\nodata  \\
\enddata
\tablenotetext{a} {In units of $12 +$ Log $N$(X)/$N$(H).}
\tablenotetext{b} {\citet{leq79}.}
\tablenotetext{c} {\citet{hid01}.}
\tablenotetext{d} {Recombination lines.}
\tablenotetext{e} {Collisionally excited lines.}
\end{deluxetable}
\clearpage

\begin{deluxetable}{lr@{$\pm$}lr@{$\pm$}lr@{$\pm$}lr@{$\pm$}lr@{$\pm$}l}
\tabletypesize{\small}
\tablecaption{NGC~6822 V, NGC~346, 30~Doradus, Orion,
and Solar Total Abundances\tablenotemark{a}
\label{tta}}
\tablewidth{0pt}
\tablehead{
\colhead{Element}  &
\multicolumn{2}{c}{NGC~6822 V\tablenotemark{b}} &
\multicolumn{2}{c}{NGC~346\tablenotemark{c}} &
\multicolumn{2}{c}{30~Doradus\tablenotemark{d}} &
\multicolumn{2}{c}{Orion\tablenotemark{e}} & 
\multicolumn{2}{c}{Sun\tablenotemark{f}}}
\startdata
12 + log He/H        & $10.909$ & 0.011  & $10.900$& 0.003   & $10.928$& 0.003   & $10.988$ & 0.003    & $10.98$ & 0.02   \\
12 + log O/H         & $ 8.42 $ & 0.06   & $ 8.15$ & 0.06    & $ 8.59$ & 0.05    & $ 8.73$  & 0.03     & $ 8.66$ & 0.05   \\
log C/O              & $-0.31 $ & 0.13   & $-0.87$ & 0.08    & $-0.45$ & 0.05    & $-0.21$  & 0.04     & $-0.27$ & 0.10   \\
log N/O              & $-1.37 $ & 0.17   & $-1.34$ & 0.15    & $-1.24$ & 0.08    & $-1.00$  & 0.10     & $-0.88$ & 0.12   \\
log Ne/O             & $-0.79 $ & 0.09   & $-0.83$ & 0.06    & $-0.76$ & 0.06    & $-0.68$  & 0.08     & $-0.82$ & 0.09   \\
log S/O              & $-1.62 $ & 0.09   & $-1.59$ & 0.12    & $-1.60$ & 0.10    & $-1.51$  & 0.05     & $-1.52$ & 0.08   \\
log Cl/O             & $-3.71 $ & 0.10   & \mcnd             & $-3.67$ & 0.12    & $-3.40$  & 0.05     & $-3.43$ & 0.06   \\
log Ar/O             & $-2.36 $ & 0.08   & $-2.33$ & 0.10    & $-2.33$ & 0.10    & $-2.11$  & 0.06     & $-2.48$ & 0.08   \\
log Fe/O             & $-1.41 $ & 0.10   & $-1.41$ & 0.10    & \mcnd             & $-1.23$  & 0.20     & $-1.21$ & 0.06   \\
\enddata
\tablenotetext{a}{Gaseous abundances for the \ion{H}{2} regions.  The
O and C abundances have been corrected for the fractions of these
elements trapped in dust grains, see text.}
\tablenotetext{b}{Gaseous abundances, values for $t^2 = 0.076 \pm 0.018$,
obtained in this paper, with the exception of the Fe/O value that
comes from stellar data \citep{ven01}.}
\tablenotetext{c}{\citet{duf82,pei00,rel02}; \citet*{pea02}, values
for $t^2$ = 0.022. The Fe/O value comes from stellar data \citep{ven99,rol03,hun05}.}
\tablenotetext{d}{\citet{pea03}, values for $t^2$ = 0.033.}
\tablenotetext{e}{\citet{cun94,est04}, values for $t^2$
= 0.024. The O and C abundances have been increased by 0.08 dex and
0.10 dex respectively to take into account the fractions of these
elements trapped in dust grains. The Cl abundance has been decreased by 0.13 dex due to an error
of +1.00 dex in the determination of the Cl$^+$/H$^+$ ratio.}
\tablenotetext{f}{\citet{chr98,asp05}.}
\end{deluxetable}
\clearpage

\begin{deluxetable}{cccccc}
\tablecaption{\ion{H}{2} region and stellar oxygen  abundances\tablenotemark{a}
\label{toxy}}
\tablewidth{0pt}
\tablehead{
\colhead{Object} &\multicolumn{2}{c}{\ion{H}{2} regions\tablenotemark{b}}
                           && \multicolumn{2}{c}{Stars}\\
\cline{2-3}
\cline{5-6}
                 &\colhead{$t^2 = 0.000$}& \colhead{$t^2 > 0.000$} 
&& \colhead{A supergiants\tablenotemark{c}} & \colhead{Sun+GCE\tablenotemark{d}}}  
\startdata
NGC 6822        & $8.16\pm0.03$ & $8.42\pm0.06$ && 8.36 & \nodata       \\
Solar vicinity  & $8.59\pm0.03$ & $8.79\pm0.05$ && 8.59 & $8.79\pm0.06$ \\
SMC             & $8.07\pm0.02$ & $8.15\pm0.04$ && 8.14 & \nodata       \\ 
WLM             &     $7.91$    &    \nodata    && 8.45 & \nodata       \\ 
\enddata
\tablenotetext{a}{In units of 12+log O/H, where it is assumed that
20\% of the O in \ion{H}{2} regions is trapped in dust grains.}
\tablenotetext{b}{The \ion{H}{2} regions are: Hubble V for NGC 6822
(this paper), the \ion{H}{2} region gradient for the Galaxy \citep{est05}, 
NGC 346 for the SMC \citep{pei00}, and the average of HM~7 and HM~9
 for WLM \citep{lee05}.}
\tablenotetext{c}{The data for NGC 6822 comes from \citet{ven01},
the data for the Galaxy and the SMC come from \citet{ven99}, 
and the data for WLM come from \citet{lee05}.}
\tablenotetext{d}{ The measured solar value is 8.66 \citep{asp05} 
which corresponds to the ISM value 4.57~Gyr ago; galactic chemical evolution models
predict an increase in O/H of 0.13~dex since the Sun was formed.}
\end{deluxetable}
\clearpage

\begin{figure}
\includegraphics[scale=.6,angle=-90]{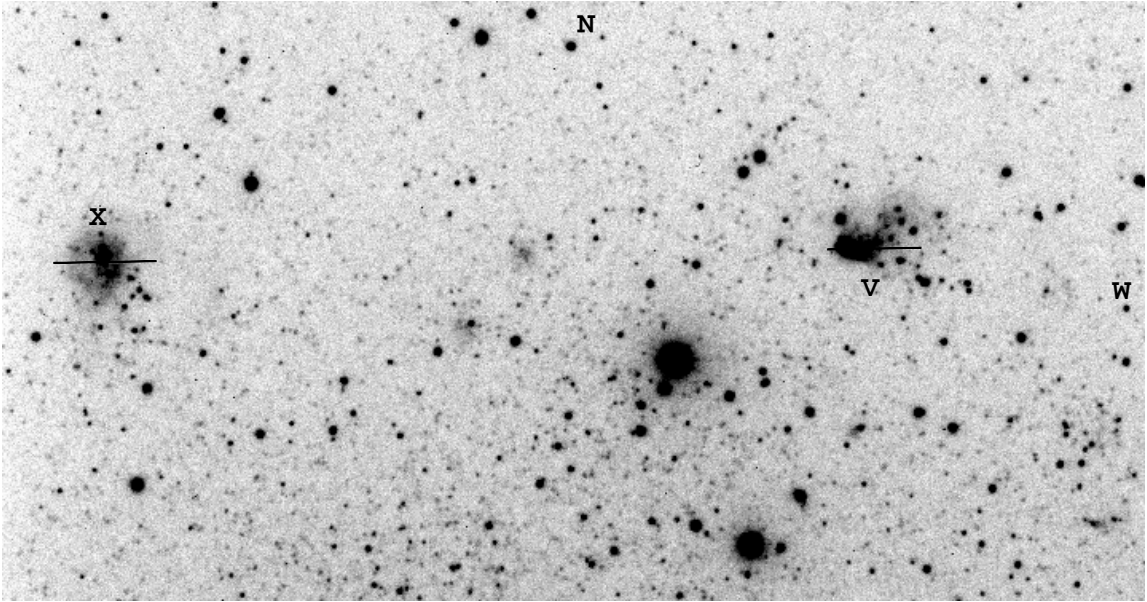}
\caption[f1.eps]{
\label{fpicture}
VLT image of regions V and X in NGC~6822. The image of
$277" \times 145"$ is centered at $\alpha = 19^{\rm h} 44^{\rm m} 57.4^{\rm s}$,
$\delta =  -14^{\rm o} 43' 26"$ (J2000), and was taken with a filter that 
suppresses light bluer than 4350~\AA. The extraction apertures 
are shown in the figure. The aperture for region V is centered 
at $\alpha = 19^{\rm h} 44^{\rm m} 52.40^{\rm s}$, $\delta = -14^{\rm o} 43' 13.4"$,
with a size of $22.4" \times 0.51"$; and the aperture
for region X is centered at $\alpha = 19^{\rm h} 45^{\rm m} 05.23^{\rm s}$, 
$\delta = -14^{\rm o} 43' 16.7"$, with a size of $24.6" \times 0.51"$.
The position angle for the long slit was $91^{\rm o}$.}
\epsscale{1.0}
\end{figure}


\begin{figure}
\includegraphics[scale=.6,angle=-90]{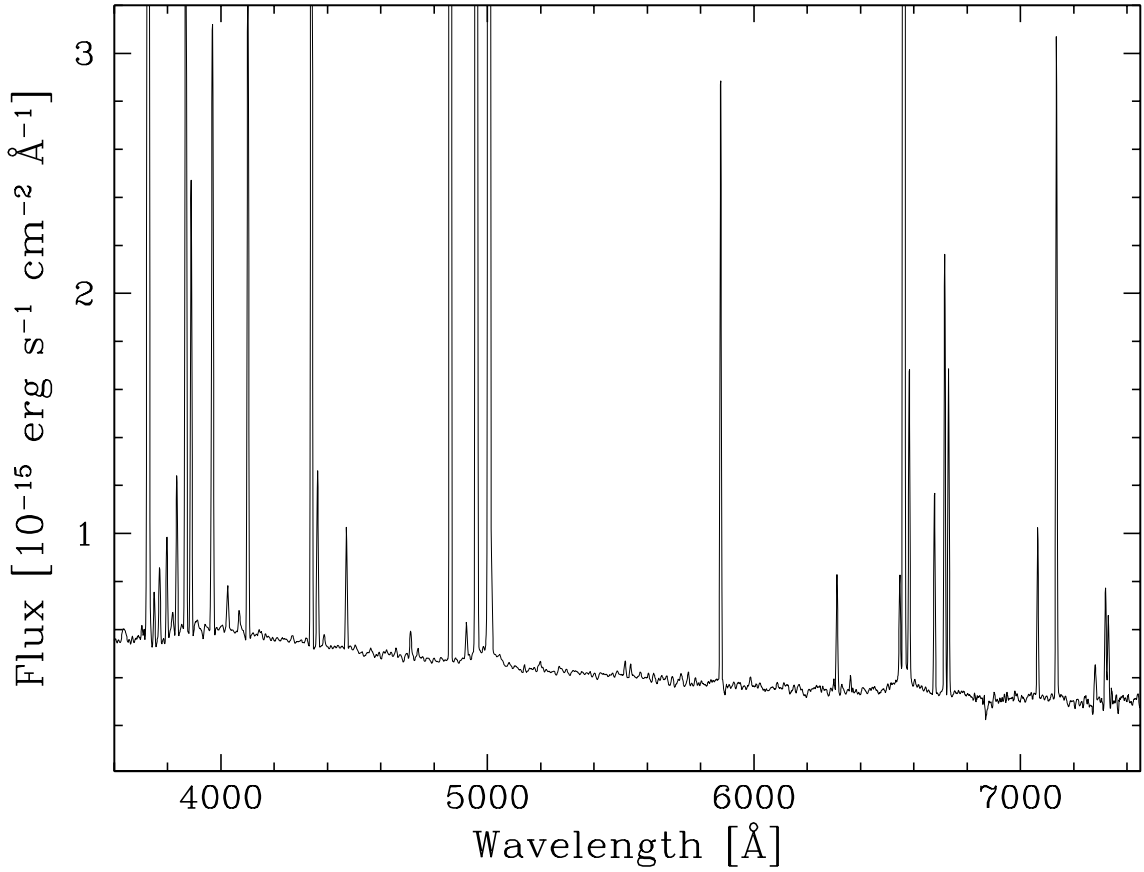}
\caption[f2.eps]{
\label{fRV}
Spectrum of region V that includes the blue and red high resolution
spectra, the spectrum has been smoothed and consequently the FWHM of the
emission lines is larger than in the raw data.}
\end{figure}


\begin{figure}
\includegraphics[scale=.6,angle=-90]{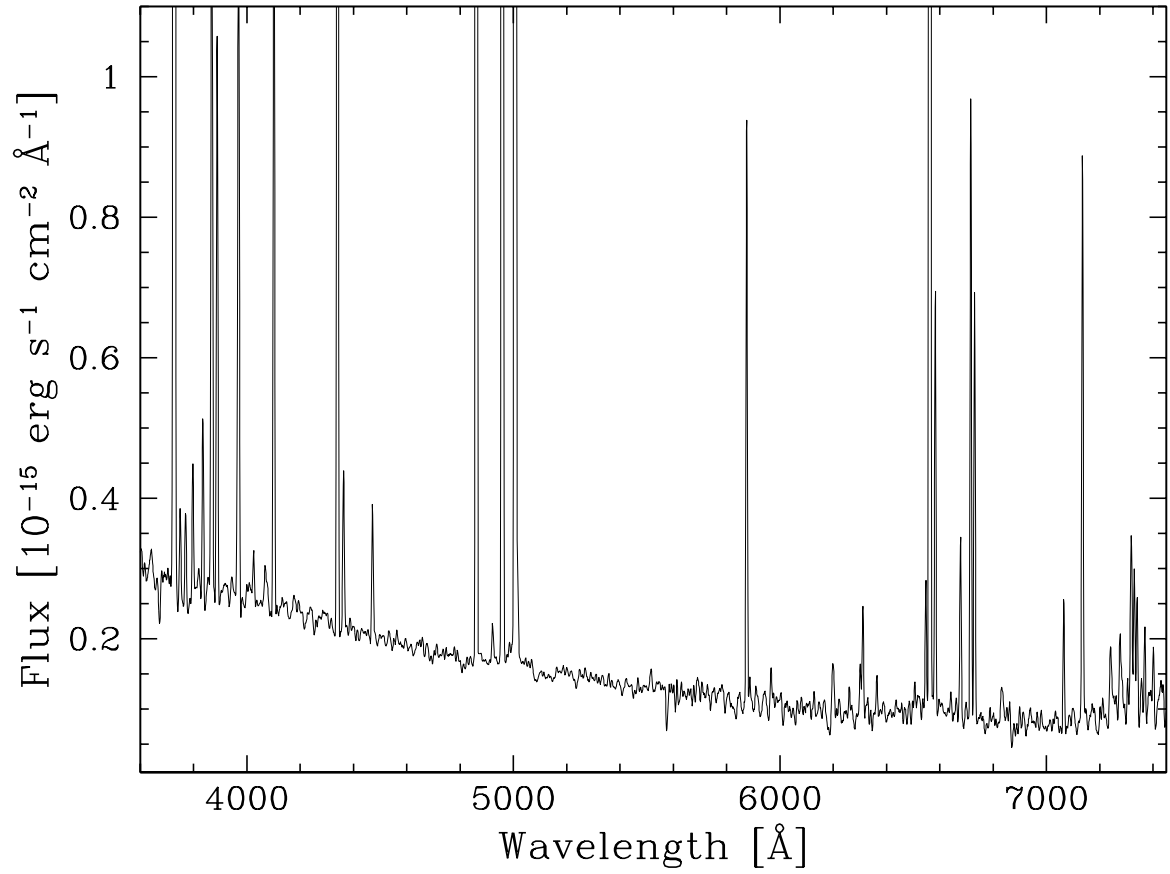}
\caption[f3.eps]{
\label{fRX}
Same as Figure~\ref{fRV}, but for region X.}
\end{figure}


\begin{figure}
\includegraphics[scale=.6,angle=-90]{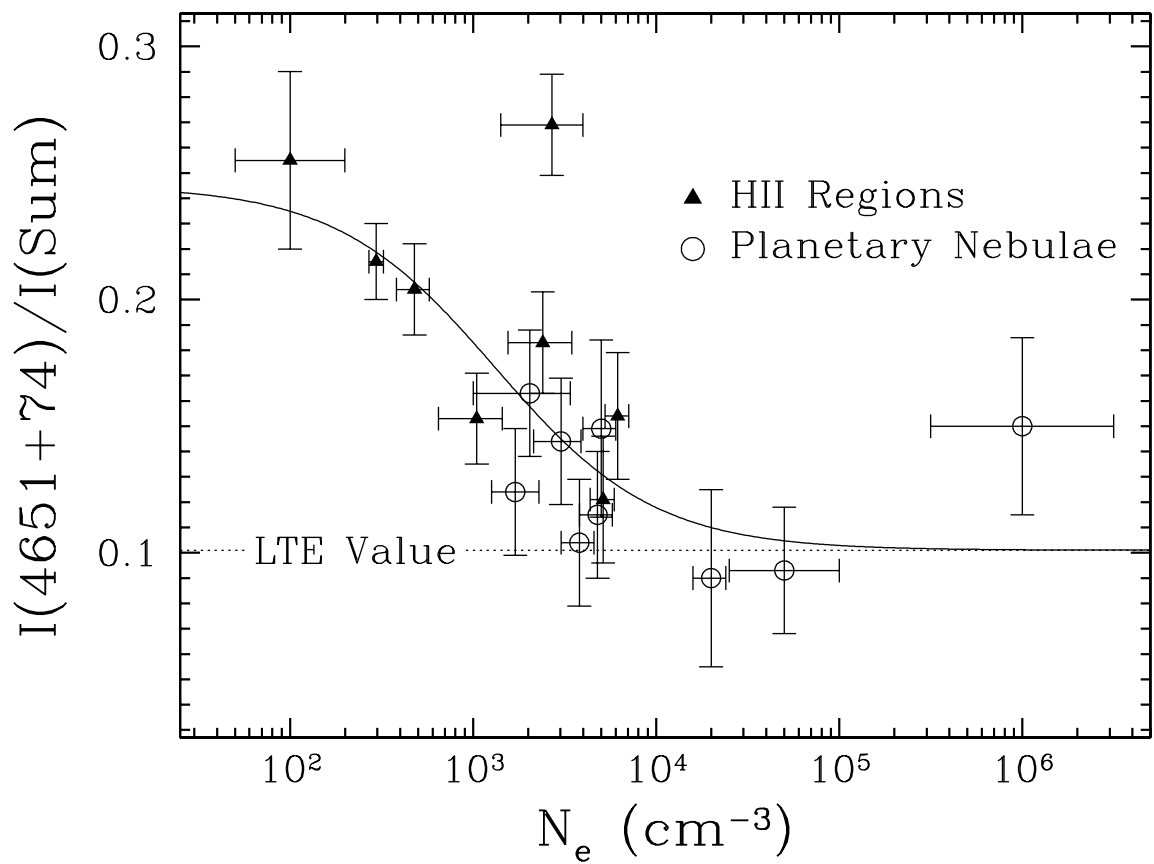}
\caption[f4.eps]{
\label{f4651}
Intensity ratio of the \ion{O}{2} multiplet 1 lines 
$\lambda$$\lambda$ 4651 and 4674 that originate in the 3p $^4$D$^0_{1/2}$ level
relative to
the sum of the intensities of the eight lines of that multiplet versus the
local electron density derived from forbidden line ratios. The solid line
represents the best fit to the data, see equation~(\ref{e4651}).}
\end{figure}


\begin{figure}
\includegraphics[scale=.6,angle=-90]{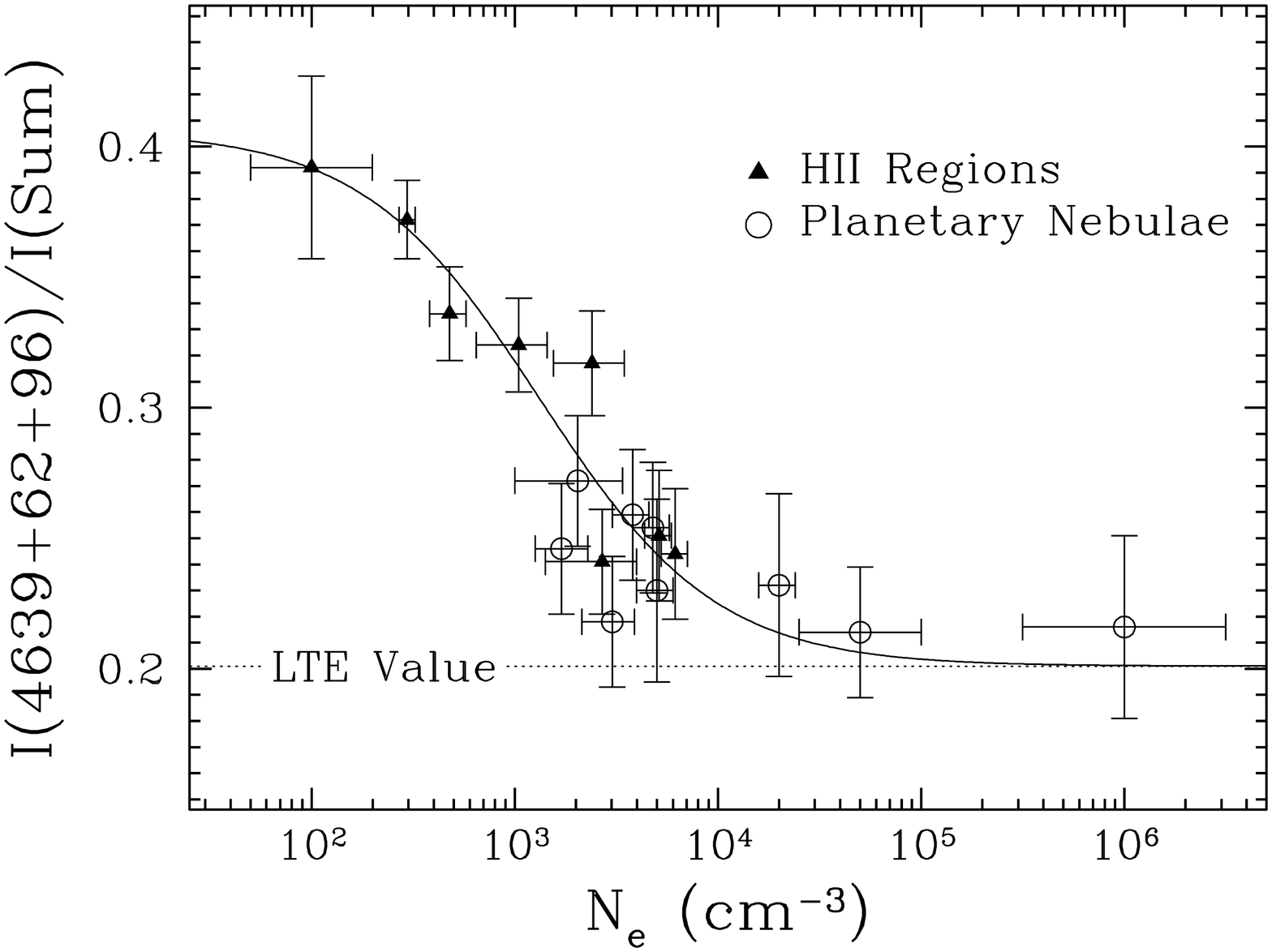}
\caption[f5.eps]{
\label{f4639}
Same as Figure~\ref{f4651}, but for 
$\lambda$$\lambda$ 4639, 4662, and 4696 that originate in the 
3p $^4$D$^0_{3/2}$ level. The solid line
represents the best fit to the data, see equation~(\ref{e4639}).} 
\end{figure}


\begin{figure}
\includegraphics[scale=.6,angle=-90]{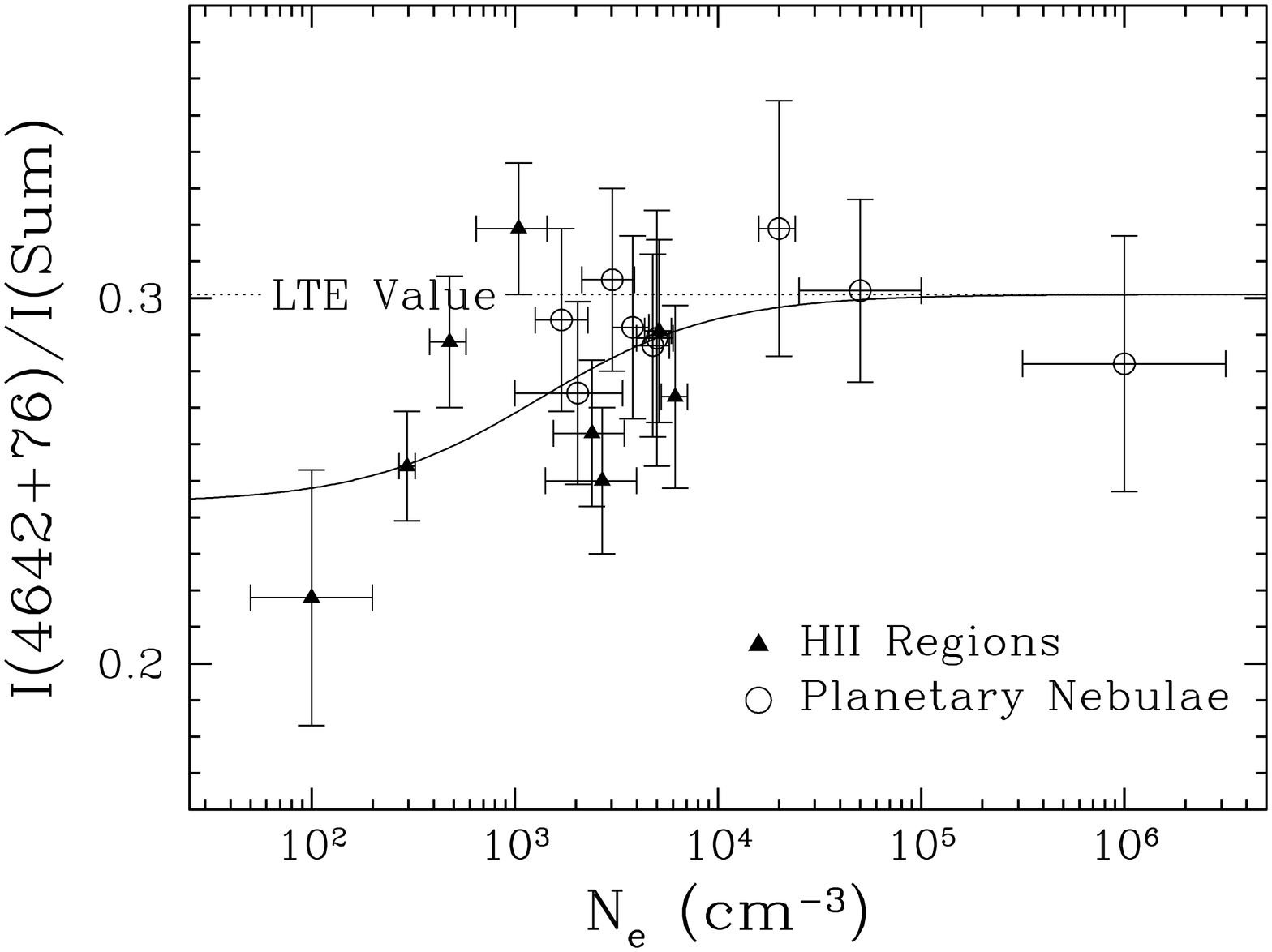}
\caption[f6.eps]{
\label{f4642}
Same as Figure~\ref{f4651}, but for
$\lambda$$\lambda$ 4642 and 4676 that originate in the 3p $^4$D$^0_{5/2}$ 
level. The solid line
represents the best fit to the data, see equation~(\ref{e4642}).}
\end{figure}


\begin{figure}
\includegraphics[scale=.6,angle=-90]{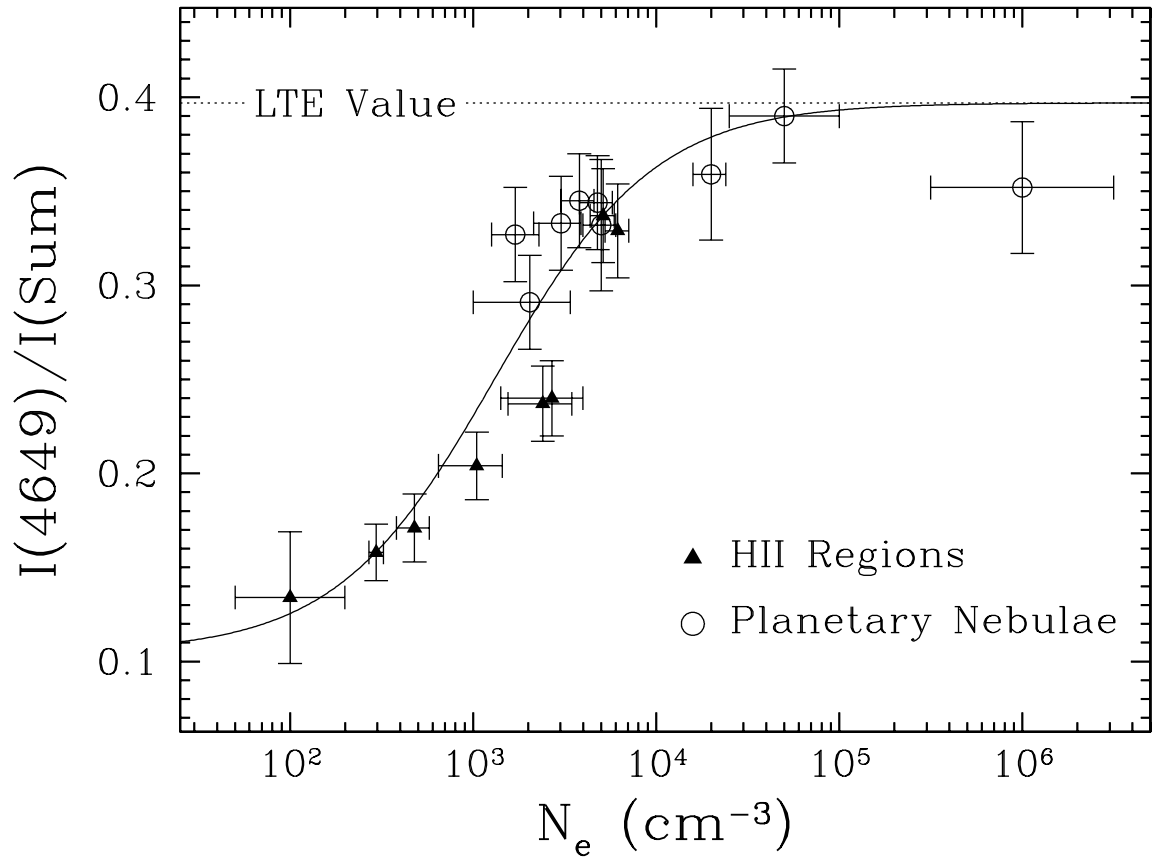}
\caption[f7.eps]{
\label{f4649}
Same as Figure~\ref{f4651}, but for
$\lambda$ 4649 that originates in the 3p $^4$D$^0_{7/2}$ 
level. The solid line
represents the best fit to the data, see equation~(\ref{e4649}).}
\end{figure}


\end{document}